\documentclass[11pt,twoside]{article}
\usepackage{./asp2014}

\aspSuppressVolSlug
\resetcounters

\begin{document}

\title{High-resolution imaging of comets}
\author{M. A. Cordiner$^{1,2}$ and C. Qi$^3$
\affil{$^1$NASA Goddard Space Flight Center, 8800 Greenbelt Road, MD 20771, USA. \email{martin.cordiner@nasa.gov}}
\affil{$^2$Institute for Astrophysics and Computational Sciences, The Catholic University of America, Washington, DC 
20064, USA.}
\affil{$^3$Harvard-Smithsonian Center for Astrophysics, 60 Garden Street, MS 42, Cambridge, MA 02138, USA \email{cqi@cfa.harvard.edu}}}

\paperauthor{Sample~Author1}{Author1Email@email.edu}{ORCID_Or_Blank}{Author1 Institution}{Author1 Department}{City}{State/Province}{Postal Code}{Country}
\paperauthor{Sample~Author2}{Author2Email@email.edu}{ORCID_Or_Blank}{Author2 Institution}{Author2 Department}{City}{State/Province}{Postal Code}{Country}
\paperauthor{Sample~Author3}{Author3Email@email.edu}{ORCID_Or_Blank}{Author3 Institution}{Author3 Department}{City}{State/Province}{Postal Code}{Country}

\begin{abstract}
Detailed mapping of the distributions and kinematics of gases in cometary comae at radio wavelengths can provide fundamental advances in our understanding of cometary activity and outgassing mechanisms. Furthermore, the measurement of molecular abundances in comets provides new insights into the chemical composition of some of the Solar System's oldest and most primitive materials. Here we investigate the opportunities for significant progress in cometary science using a very large radio interferometer. The ngVLA concept will enable detection and mapping of a range of key coma species in the 1.2-116~GHz range, and will allow for the first time, high-resolution mapping of the fundamental cometary molecules OH and NH$_3$. The extremely high angular resolution and continuum sensitivity of the proposed ngVLA will also allow the possibility of imaging thermal emission from the nucleus itself, as well as large dust/ice grains in the comae, of comets passing within $\sim1$~au of Earth.
\end{abstract}

\section{Comets: messengers from the birth of the Solar System}

Comets are thought to have accreted in the Solar System at around the same time as the planets. Depending on the degree of subsequent thermal and radiative processing, cometary nuclei may contain pristine material from the proto-Solar disk or prior interstellar cloud. Studies of cometary ices therefore provide unique information on the physical and chemical properties of the early Solar System and pre-solar Nebula \citep{mum11}. By virtue of their organic-rich composition, cometary impacts could have been important for initiating prebiotic chemistry on the early Earth, and examining their molecular content provides crucial details on the relationship between interstellar and planetary material \citep{ehr00,ale18}.

To a first-order approximation, the structure of cometary comae can be interpreted using a uniform, spherically-symmetric outflow model, which defines volatile molecules as either `primary' (parent) species, sublimated directly from the nucleus or photo-chemical `product' (daughter) species produced in the coma \citep{has57}. Molecules commonly detected in the coma using ground-based telescopes include H$_2$O, CO, CH$_4$, CH$_3$OH, NH$_3$, HCN, C$_2$H$_2$, C$_2$H$_6$, H$_2$S and CS but detections of more complex molecules such as HC$_3$N, as well as above-average abundances of HNC, H$_2$CO, CN, C$_2$ and C$_3$ in some comets imply that other organic parents are likely present \citep{cro04,cot08,lis08}, probably in the form of organic polymers or macro-molecules.  Distributed coma sources are known or suspected for several commonly-observed species (including CN, HNC and H$_2$CO), originating from presently unknown precursor material(s) at distances up to $\sim10^5$~km from the nucleus \citep{biv99,cor14,cor17}. The clearest measure of release from the nucleus (\emph{vs.} production in the coma) requires measurement of the detailed spatial distribution of a species, especially its variation with nucleocentric distance in the innermost coma, within a few hundred to a few thousand km of the nucleus.

\section{Elucidating the origins of cometary gases through radio interferometry}

A fundamental goal of cometary  investigations is to understand the origin and nature of cometary nuclei, in order to explore the chemical link between mature planetary systems and the material present in the accretion disk from which they formed. Of particular interest are the primary volatiles, as measurement of their production can reveal relative abundances of species originally incorporated into
a particular cometary nucleus, and therefore information about the conditions under which the comet formed. 

The Rosetta mission to comet 67P demonstrated unequivocally the complex nature of cometary outgassing and activity. Strong chemical differentiation was observed for outgassing from different parts of the nucleus \citep{boc16}, and narrow, highly directional jets were observed during outburst events as the comet approached perihelion (see Figure \ref{fig:jets}). Such features typically span kilometer (milliarcsecond) scales, which can only be resolved from the ground using the high-sensitivity, long-baseline interferometry.

\articlefigure{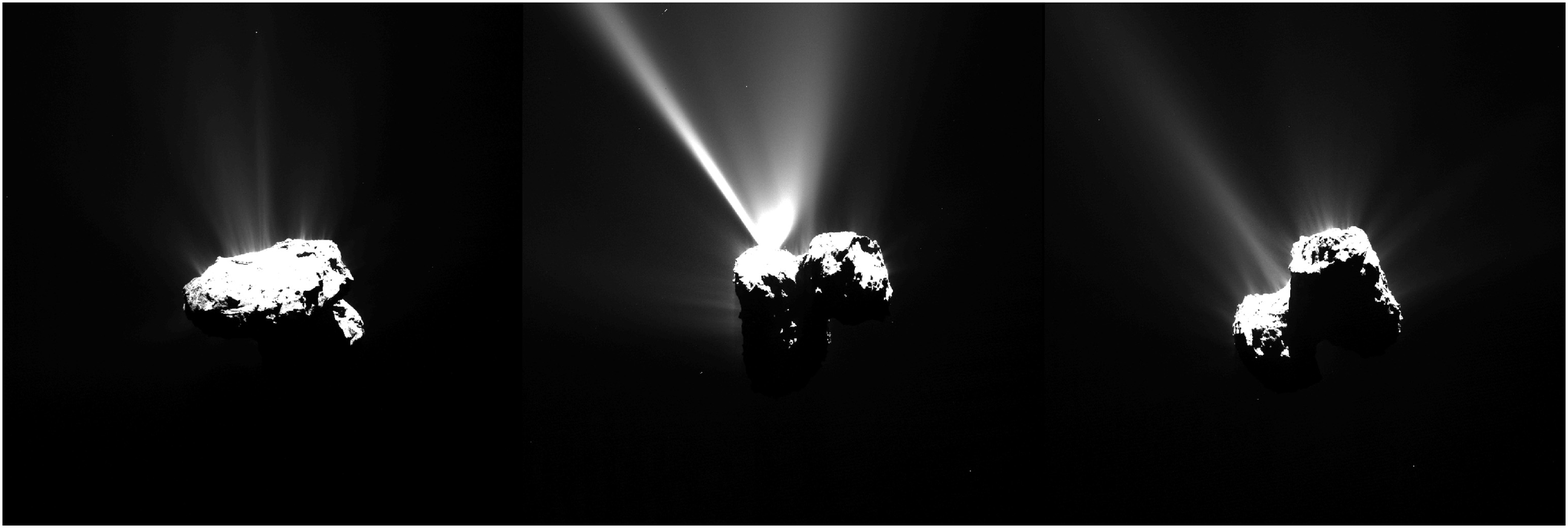}{fig:jets}{Optical images of comet 67P/Churyumov-Gerasimenko from Rosetta's OSIRIS camera, obtained near perihelion on 12 August 2015, showing strong, narrow jets of gas and dust. Credit: ESA/Rosetta/MPS for OSIRIS Team MPS/UPD/LAM/IAA/SSO/INTA/UPM/DASP/IDA.}

Radio interferometers can map the spatial distributions of multiple primordial molecules released directly from the interior of the nucleus. Heterodyne receiver technology enables the derivation of high-resolution line-of-sight velocity information, from which 3-D structures can be obtained for the observed species, assisting in investigations of the chemical heterogeneity and true composition of the nucleus \citep{qi15}. The first reported interferometric observation of a comet was of the 18 cm OH line in 1P/Halley, detected and mapped using the VLA by \citet{dep86}. Formaldehyde (H$_2$CO) was then detected with the VLA by \citet{sny90} in comet Machholz, and subsequently mapped in comet Hale-Bopp by \citet{mil06} using the BIMA interferometer. 
Figure \ref{fig:ovro} shows OVRO observations of molecular emission from the inner coma of comet Hale-Bopp \citep{bla99},
in which the HNC, DCN, and HDO are seen to lie in arcs or partial shells, offset from the nucleus of the comet. This was likely caused
by asymmetric outgassing/jet activity, which lifted a population of icy grains from the nucleus that sublimated to release the observed gases. Interferometric mapping is thus required to properly elucidate the production mechanisms of cometary species, to enable the derivation of accurate mixing ratios.

\articlefigure{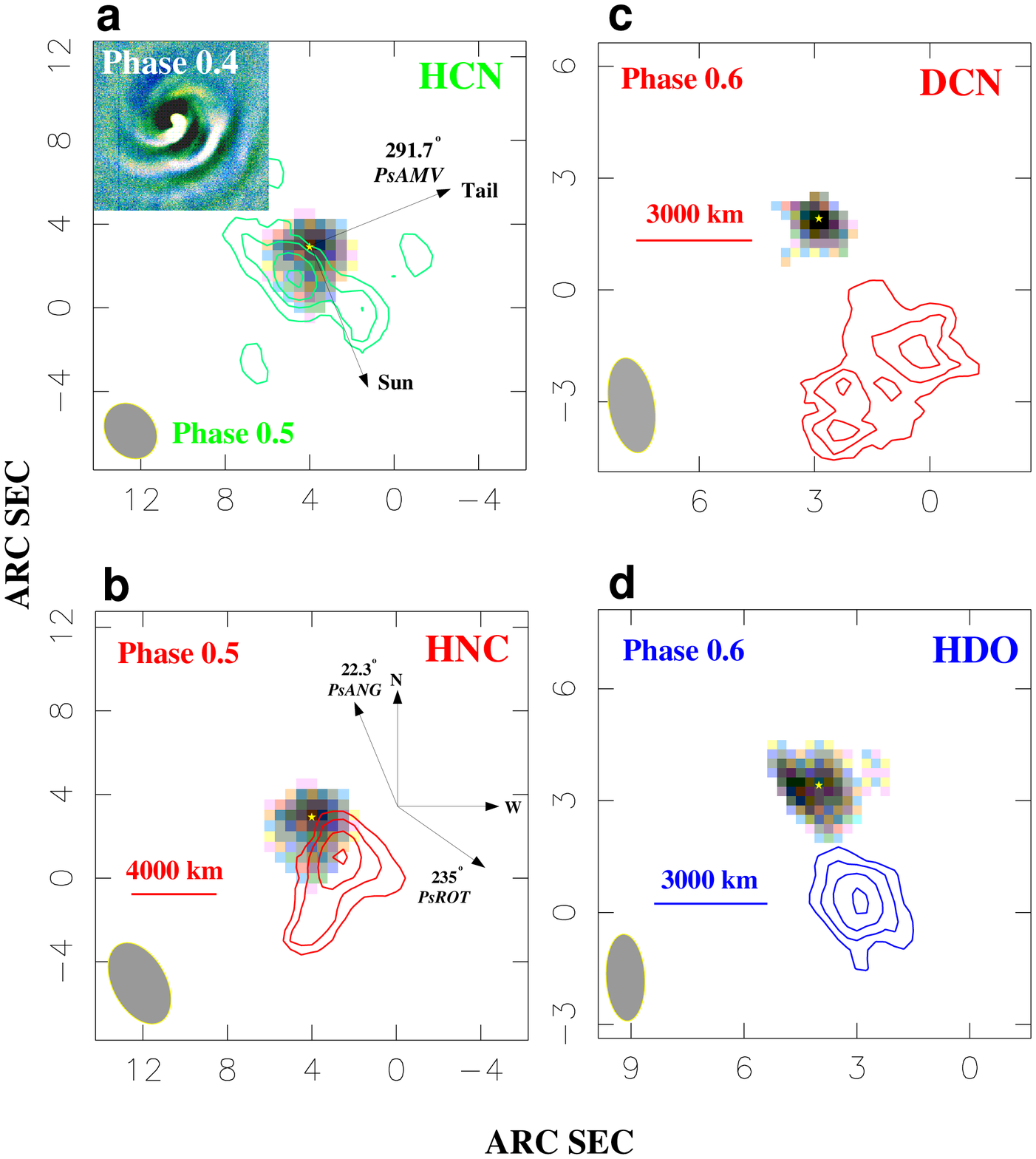}{fig:ovro}{OVRO 3~mm observations of molecular emission from the inner coma of comet Hale-Bopp \citep{bla99}. The scattered light optical image (upper left) is dominated by a precessing nuclear dust jet. The star marks the position of the nucleus, and the grey scale presents the continuum emission. Contours in the various panels depict molecular emission from the species listed in the upper right corner.  Arrows show the direction of the sun, the cometary rotation axis, and the motion of the comet on the sky. Note that the HCN and DCN emissions were observed on different days but are compared at the same rotational phase.}

The importance of radio interferometry to distinguish between molecules arising from different parts of the nucleus and coma was confirmed by \citet{cor14,cor17}. Observations of two comets (C/2012 S1 (ISON) and C/2012 F6 (Lemmon)) were made using the Atacama Large Millimeter/submillimeter Array (ALMA) at frequencies 339-365~GHz (0.82-0.89~mm) with baselines 15-2700~m, resulting in an angular resolution of approximately $0.5''$. Figure \ref{fig:alma} shows spectrally-integrated maps for three different molecules (HCN, HNC and H$_2$CO) in these two comets, and dramatic differences between their spatial distributions are evident. 

Modeling of the interferometric visibility amplitudes \citep[\emph{e.g.}][]{boi07} enables the coma origins of the observed species to be derived. In both ISON and Lemmon, HCN was found to originate from (or very close to) the nucleus, with a spatial distribution largely consistent with spherically-symmetric, uniform outflow. The HNC and H$_2$CO distributions, on the other hand, were found to be consistent with release as products of coma chemistry --- \emph{i.e.} they do not originate directly from the nucleus, but most likely arise from the breakdown of large organic molecules/polymers in the coma. The observed molecular distributions were found to be variable on a timescale of days to minutes \citep{cor17}. Thus, the importance of `snapshot' interferometry of cometary comae is demonstrated.

The proposed ngVLA concept will provide important improvements for cometary studies compared with currently available facilities. In particular, the unique centimeter-wave capabilities will open up the possibility of high-resolution OH and NH$_3$ mapping for the first time. The exceptional $uv$ coverage and surface brightness sensitivity of ngVLA will provide an order of magnitude improvement in signal-to-noise for mapping these species compared with existing state-of-the-art interferometers such as the JVLA.

\articlefigure{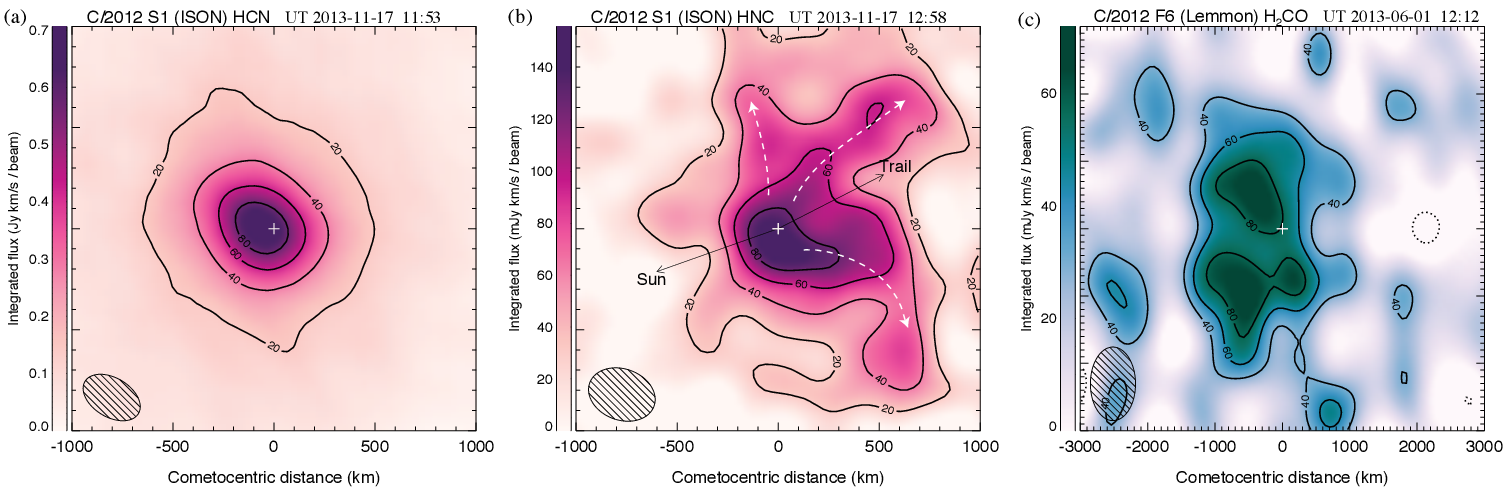}{fig:alma}{Contour maps of spectrally-integrated molecular line flux observed in comets S1/ISON and F6/Lemmon using ALMA. Contour intervals in each map are 20\% of the peak flux (the 20\% contour has been omitted from panel (c) for clarity). On panel (b), white dashed arrows indicate HNC streams/jets.  The peak position of the (simultaneously observed) 0.9~mm continuum is indicated with a white `$+$'. See \citet{cor14} for further details.}

\section{Measuring the chemical composition of the nucleus}

While the chemistry of interstellar clouds and collapsing protostellar envelopes is known from radio and IR spectroscopy \citep{her09}, our knowledge of protoplanetary disk compositions is comparatively sparse \citep{wal16}. Theoretical models show that disk chemistry is very sensitive to the thermal, radiative and dynamical history \citep{dro16}, which are not well constrained for the protosolar nebula.  Understanding the chemical evolution of matter as it passes from interstellar to planetary phases is crucial for theories concerning the accretion and composition of planetary bodies and their atmospheres, but this work is hindered by the extreme difficulty of disk mid-plane ice (and gas) observations.  Measurements of cometary compositions provide a unique method for probing the chemistry of the protosolar disk mid-plane, and may be used to test theories regarding the chemical evolution of the early solar system.

To determine the capabilities of the proposed ngVLA concept with regard to cometary composition studies, we generated spectral line flux simulations for a range of commonly-observed coma molecules. These simulations are based on a spherically-symmetric (Haser-type) outflow model in LTE at 75~K, for a typical (moderately bright) comet at 1 au from Earth, with a water production rate $Q({\rm H_2O})=2\times10^{29}$~s$^{-1}$ and outflow velocity 0.8~km\,s$^{-1}$. Assumed abundances (with respect to the dominant volatile H$_2$O) are based on typically-observed values \citep{mum11}.  The coma morphology is dominated by a 1/$\rho$ brightness distribution (where $\rho$ is the sky-projected distance), and our calculated fluxes are for a nucleus-centered ngVLA beam, tapered to a resolution of $1''$.

Results of these calculations are given for the strongest line of each species (in the nominal ngVLA frequency range 1.2-116~GHz) in Table 1; for many species, additional lines will be observable. We have only included molecules for which detections will be readily achievable based on current estimates for the ngVLA sensitivity (with line fluxes $\gtrsim1$~mJy per $1''$ beam per 1.6~km\,s$^{-1}$ unit bandwidth), assuming 1 hour on-source. Accurate, instantaneous OH and NH$_3$ abundance measurements will thus be possible at spectral line sensitivities 50-100~$\mu$Jy, which are expected to be achievable using ngVLA, but would take weeks of integration with the JVLA. In addition to these fundamental cometary species, other species such as sulphur-bearing molecules and more complex organics may also be detectable, permitting detailed studies of cometary compositions beyond what is possible using currently-available instruments.

\begin{table}
\caption{Calculated fluxes for representative cometary lines using ngVLA concept ($1''$ beam)}
\begin{center}
\begin{tabular}{llccc}
\hline\hline
Species&Transition&Frequency&Abund.&Flux\\
&&(MHz)&(\%)&(mJy)\\
\hline
OH&$J=3/2$, $F=2-2$&1667&86&1.7\\
NH$_3$&$J_K=3_3-3_{-3}$&23870&1.0&1.0\\
H$_2$CO&$J_{K_aK_c}=1_{01}-0_{00}$&72838&0.5&1.7\\
HCN&$J=1-0$&88632&0.2&13.9\\
HNC&$J=1-0$&90664&0.02&2.7\\
HC$_3$N&$J=10-9$&90979&0.02&3.0\\
CS&$J=2-1$&97981&0.1&7.2\\
CH$_3$OH&$J_K=3_1-4_0\, A^+$&107014&2.0&4.9\\
CH$_3$CN&$J_K=6_0-5_0$&110383&0.02&1.5\\
CO&$J=1-0$&115271&5.0&2.3\\

\hline
\end{tabular}
\\
\parbox{0.8\textwidth}{\noindent \footnotesize Flux calculations are based on a Haser parent model for a typical comet at 1 au (see text). For OH, the predicted flux was scaled from the VLA observation of comet Wilson by \citet{pal89}, with a representative inversion parameter of 0.4.}
\end{center}
\end{table}

\section{Revealing the coma structure through high-resolution molecular imaging}

Cometary activity occurs as a result of solar radiation that heats the volatile ices in the nucleus as the comet approaches the inner solar system. Despite decades of remote and in-situ studies, the physical outgassing mechanisms and primary drivers of cometary activity remain to be fully understood \citep{gun15}.

The unique capability of ngVLA to map OH and NH$_3$ at high spatial resolution (see Section 2), will allow a leap forward in our ability to probe the detailed outgassing behaviour of comets. Hydroxyl (OH) and ammonia (NH$_3$) are among the most abundant cometary molecules, but their centimeter-wave transitions have never before been mapped in a cometary coma at high ($\sim1''$) resolution. The spatial distribution of NH$_3$ in the coma is largely unknown, so detailed mapping is expected to reveal important new insights into the nature of the main reservoir of cometary nitrogen.  Ammonia  will be detectable in the proposed ngVLA receiver band 4 through its inversion transitions around 23.7-23.9~GHz, whereas OH has strong fine structure transitions around 1.7 GHz. Release of NH$_3$ from the nucleus may be mapped in active comets within a few astronomical units of Earth, allowing the detailed structure of the inner coma to be measured on scales $\sim1''$ ($\sim1000$~km). This will reveal jets and other outgassing features, similar to those detectable by ALMA in other molecules (Figure \ref{fig:alma}).  By contrast, OH is produced further from the nucleus as a result of H$_2$O photolysis, and can be used to map the larger-scale coma structure \citep[\emph{e.g.}][]{dep86}. To fully capitalize on the unique sensitivity and mapping capability of ngVLA for comets, the need for long ($\sim100$~km) as well as short ($\sim50$~m) baselines is emphasized, to allow measurement of the coma structure on (milli-)arcsecond to arcminute scales. Cometary comae tend to be centrally peaked (with a $\sim1/\rho$ brightness distribution), but can extend to over $100''$, so our science case would also benefit from high-sensitivity, simultaneous single-dish observations to fill in lost flux from the largest coma scales.

Resolved spectral line profiles for the observed gases (at $\sim0.1$~km\,s$^{-1}$ resolution) will reveal the detailed coma kinematics, and the gas temperatures may be obtained by measuring multiple NH$_3$ lines from the different fine structure levels of this molecule. This will enable new tests for our understanding of coma physics.

\section{Resolving the nucleus using long-baseline interferometry}

Cometary nuclei are among the most difficult objects of the solar system to detect and characterize \citep{lam04}. Rendezvous missions such as Giotto, Deep Impact and Rosetta have provided our only close-up views of a handful cometary nuclei, revealing diverse morphologies and surface properties that provide information on their origins as well as thermal and collisional histories. Direct imaging of cometary nuclei using traditional ground-based observing strategies is extremely difficult due to their faintness and small sizes (typically $\lesssim10$~km, or $14$ mas at 1 au). In addition, dust emission (and reflected sunlight) tends to swamp the weak signal from the nucleus in the optical and infrared. Although thermal emission from large dust grains in the coma contributes significantly to the cometary continuum signal in the centimeter and millimeter bands, the much broader extent of the coma dust emission allows it to be identified and separated from the compact nucleus contribution.

Detections of nucleus thermal emission have previously been possible only in the case of rare, extremely favourable cometary apparitions \citep[\emph{e.g.}][]{boi14}. The ngVLA will provide, for the first time, the ability to routinely observe the nuclei of comets and other near-Earth objects. Assuming an albedo of 0.05, the (black body) continuum flux from a 5~km-diameter comet within 0.5 au of Earth will be easily detectable at around 0.4 mJy. Observed with ngVLA at an angular resolution of 2 mas at 115 GHz (assuming 300~km baselines), the nucleus of such a comet would be well-resolved (with 7 resolution elements across), allowing accurate size and shape measurements. 

\section{Conclusion}

Due to their relatively low surface brightness, cometary comae are extremely challenging to image at radio wavelengths. Only since the advent of ALMA has spectral mapping of moderately bright comets become routine, resulting in a paradigm shift in our understanding of the chemistry and physics of the molecular coma. However, the limited frequency coverage of ALMA prohibits the study of fundamental cometary molecules such as OH and NH$_3$, which will be uniquely accessible for high resolution mapping by ngVLA. Furthermore, in this article we have demonstrated the possible power of ngVLA for detecting and mapping a range of key coma species in the 1.2-116~GHz range, as well as the possibility of detecting dust, and thermal emission from the nucleus itself. Interferometry of comets with sub-millijansky sensitivity for narrow lines in the centimeter-millimeter spectral region will enable new insights into the composition of the nucleus, and consequently, the early history of our Solar System. 

\acknowledgements This work was supported by the National Science Foundation under Grant No. AST-1614471.



\end{document}